\documentclass[10pt,twocolumn,letterpaper]{article}
\usepackage[pdftex]{graphicx}
\usepackage[margin=0.5in]{geometry}
\usepackage[font=scriptsize]{caption}
\usepackage[affil-it]{authblk}

\begin{document}
\title{Optimization of Frequency Selective Surface Bandstop Filter Based on Fractal Unit Cell}
\author{Jordon Fletcher}
\affil{Columbia University}
\date{}

\twocolumn[
  \begin{@twocolumnfalse}
    \maketitle
    \begin{abstract}
      In this paper, a Frequency Selective Surface fractalized unit cell design previously explored in the literature is replicated then extended and optimized for multilayer FSS morphologies.  A 10 mm x 10 mm planar unit cell is used and then iterated on to create a multilayer structure optimized for use as a bandstop filter in the X-band and Ku-band.  Substrate thickness, unit cell shift, and number of layers are varied and numerical results are obtained via the finite element method employed by Ansys solver for electromagnetic structures (HFSS).  S-parameters for reflection and transmission are obtained at various stages of design optimization to verify the desired bandstop results, and bandpass characteristics are explored at each stage of the design as well.

\hfill \break

    \end{abstract}
  \end{@twocolumnfalse}
]

\section{Introduction}

The proliferation of space and satellite communication links have made the X and Ku bands an attractive portion of the radio spectrum for both researchers and industry.  Numerous applications exist within these bands, including radar and high-resolution imaging, wireless computer networks, satellite television, and deep space communications [1].  Frequency selective surfaces (FSS) have found usage in many areas of communications technology, and are often employed as bandstop or bandpass filters, particularly in the X and Ku bands [2, 3].  Typically thin, repetitive arrangements of metallic elements on a dielectric substrate, FSS consist of a unit cell repeated in a planar fashion and may also extend into three dimensions via multiple layers.  An alternative FSS type consists of the unit cell geometry etched into a metallic ground plane.  They have been used in many applications including antenna design, satellite broadcasting, and reduction of radar cross section (RCS) for military vehicles [4].    FSS are often characterized as first, second, or higher order depending on what order bandpass or bandstop response they exhibit [5].  Typically to achieve higher-order responses multilayer FSS are employed.  A number of factors are important when designing FSS for specific transmission and reflection properties, including element type and geometry, spacing, substrate thickness, and inter-element spacing [6, 7].  

In [8], the authors sought to create a single-layer FSS which possessed bandstop filter properties in both the X and Ku bands.  The novel unit cell geometry utilized consisted of an evolving fractal shape of branching arms orthogonal to a central trunk (Fig. 1).  The authors focused on variations in arm length and the accompanying bandstop frequency changes.  In this paper, this work is expanded upon with the addition of multiple layers using the fractal unit cell geometry and variations in substrate thickness.  The work in [8] is first replicated to verify results, then additional layers are added and substrate thickness along with pattern offset is parametrically swept to obtain both transmission and reflection results.  

\section{Single Layer FSS}
The geometry of the unit cell employed by [8] consisted of four stages evolved over variations in arm length to achieve the final fractalized result.  Here the final design consisting of all four stages depicted in Fig. 1 was taken as the initial starting point.

\begin{figure}[h]
\center
    \includegraphics[width=0.7\columnwidth]{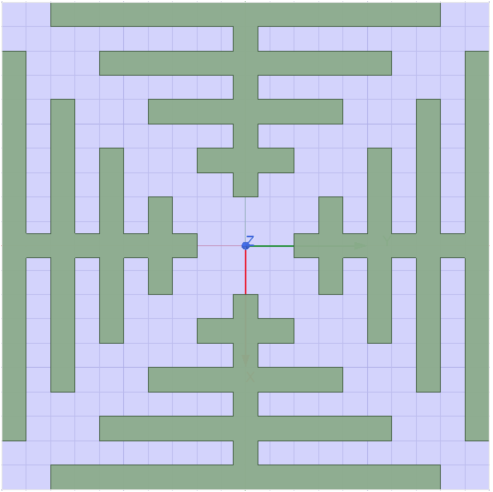}
    \caption{Four arm fractal FSS unit cell}
\end{figure}

The total unit cell dimensions were 10 mm by 10 mm, on a FR-4 (Fire Resistant-4) substrate with relative permittivity $\epsilon_r = 4.4$, loss tangent = 0.2, and thickness of 0.7 mm.  With these conditions transmission characteristics were obtained with good agreement of the results reported by [8] as shown in Fig. 2.  

\break

\begin{figure}[h]
\center
  \includegraphics[width=\linewidth]{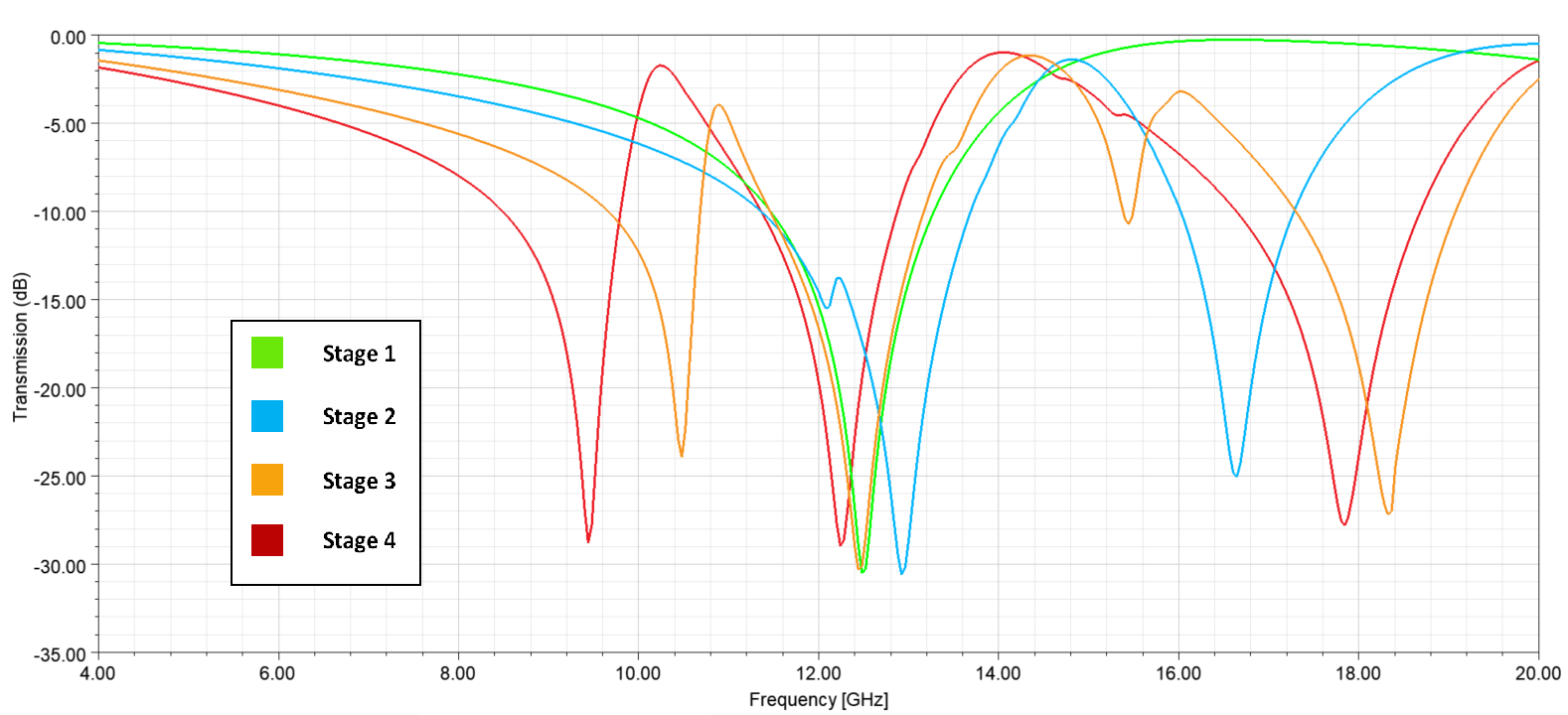}
  \caption{Transmission Coefficients of four evolving unit cell stages}
\end{figure}

\begin{figure}[h]
\center
   \includegraphics[width=\linewidth]{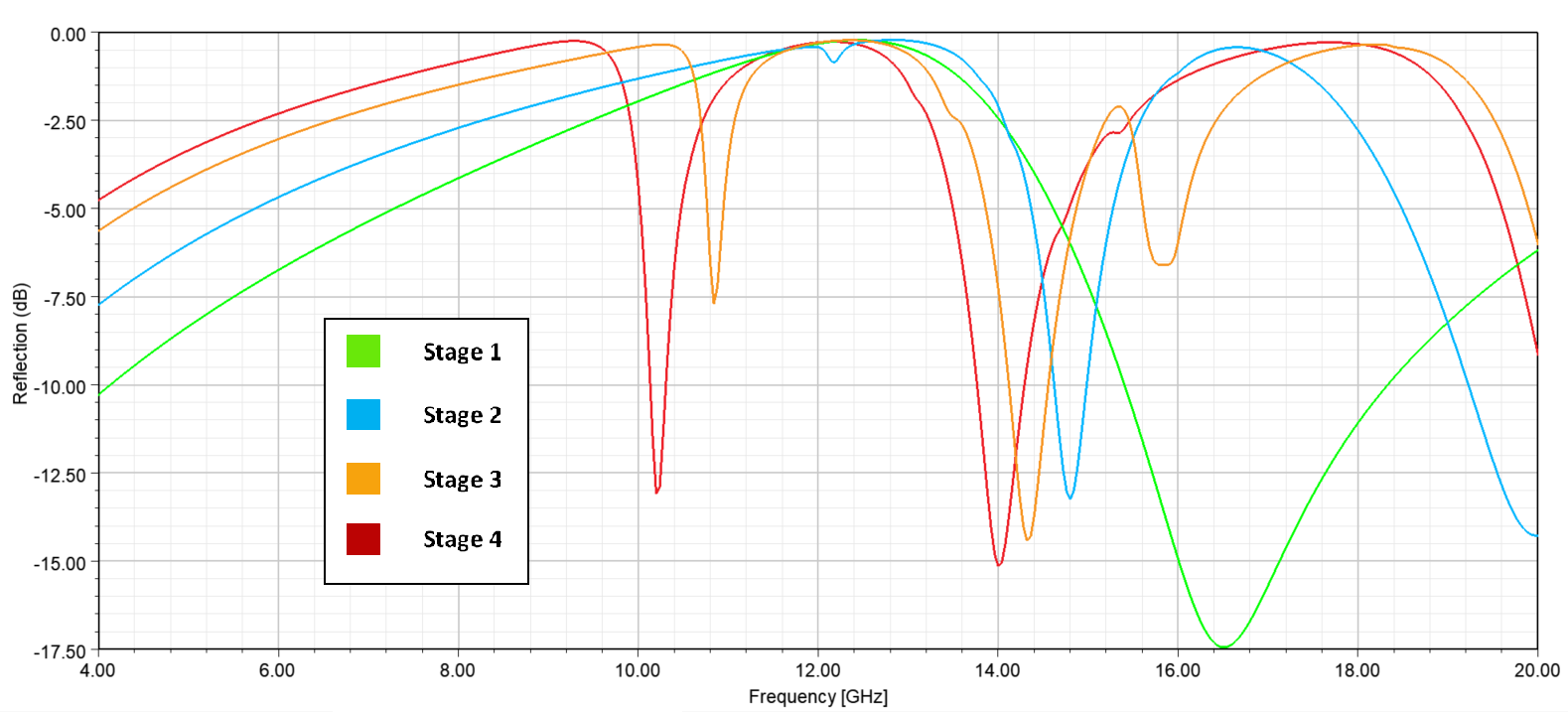}
    \caption{Reflection coefficients of four evolving unit cell stages}
\end{figure}

The authors reported structure resonances at 8.6 GHz, 11.9 GHz, and 17.5 GHz for the final four-stage design with bandwidths of 1.6 GHz, 1.4 GHz, and 1.9 GHz respectively.  Results obtained here show similar resonances for the same design at 9.4 GHz, 12.2 GHz, and 17.8 GHz with bandwidths of 1.3 GHz, 1.5 GHz, and 2.0 GHz measured at the -10 dB point.  No reflection characteristics were reported in [8], but here reflection coefficient data was obtained and is shown in Fig. 3.  The data shows the FSS structure exhibits bandpass characteristics at 10.2 GHz and 14.0 GHz with return loss values of -13.1 dB and -15.1 dB respectively.  

\section{Two-Layer FSS Parameter Variation and Results}

After confirming model predictions against prior results, the FSS unit cell and substrate were duplicated to create a second layer (Fig. 4).  

\begin{figure}[h]
\center
   \includegraphics[width=\linewidth]{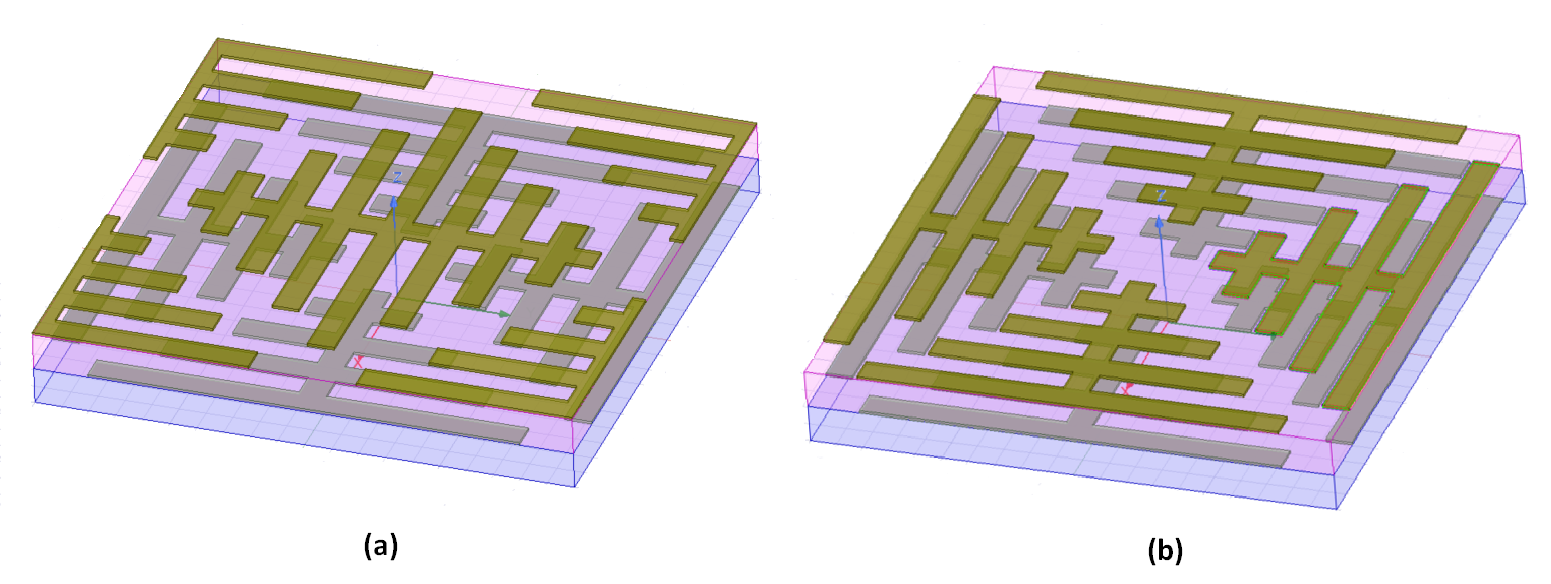}
    \caption{Two-layer FSS with a) 5 mm (max) offset of the top layer pattern and b) 0 mm offset}
\end{figure}

Top layer pattern offset relative to the bottom layer was varied to determine impact on transmission and reflection parameters.  For the case of zero offset, results showed three bandpass regions at 6.1 GHz, 9.6 GHz, and 13.9 GHz (Fig. 5).  A bandpass region also appears to exist around 20 GHz though this was outside the range of calculated frequencies and so is not reported.

\begin{figure}[h]
\center
   \includegraphics[width=\linewidth]{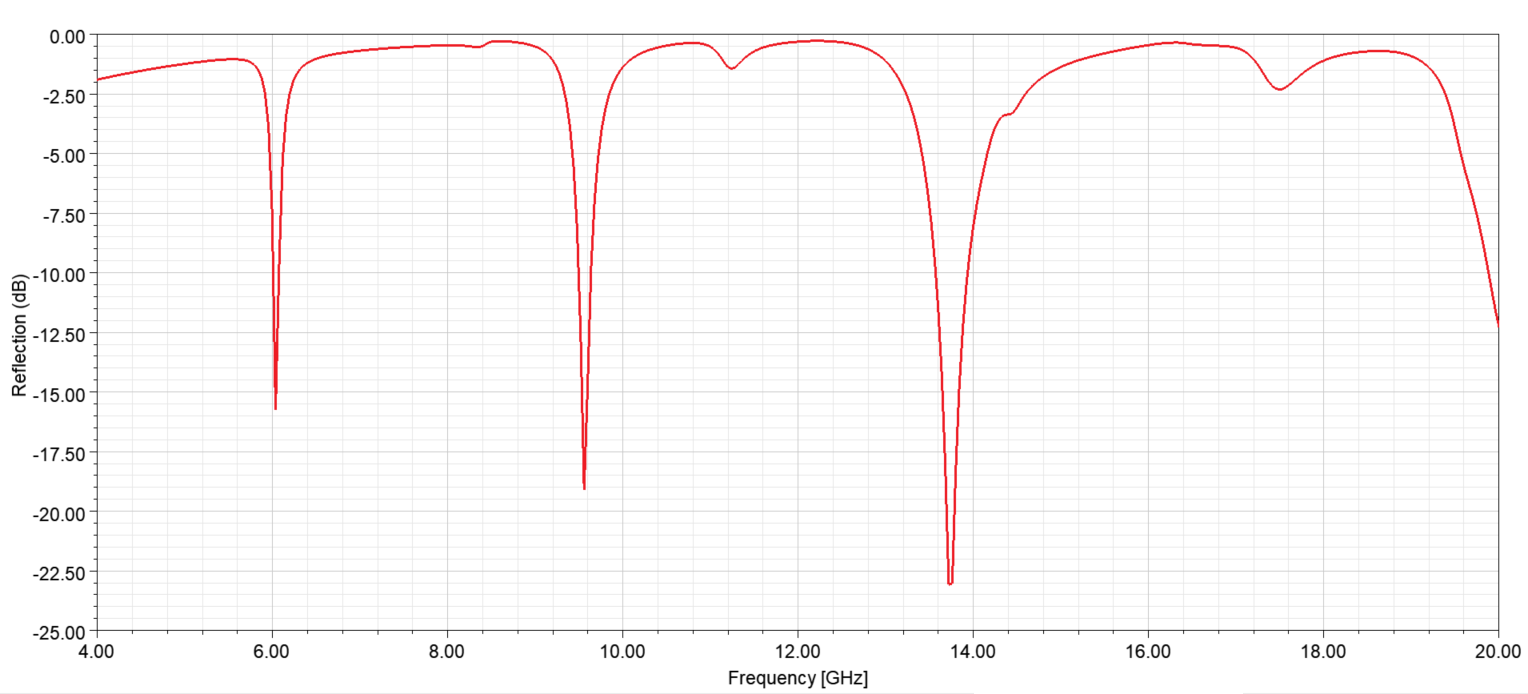}
    \caption{Reflection coefficient of two-layer FSS with 0 mm offset of the top pattern}
\end{figure}

Bandstop features as evidenced by transmission characteristics (Fig. 6) were more complex than the single layer case, with multiple apparent bandstop regions across the frequency range of interest.  Primary structure resonances with transmission at -20 dB or lower occurred at 9.2 GHz, 11.0 GHz, 12.5 GHz, and 19.2 GHz (Fig. 6).  

\begin{figure}[h]
\center
   \includegraphics[width=\linewidth]{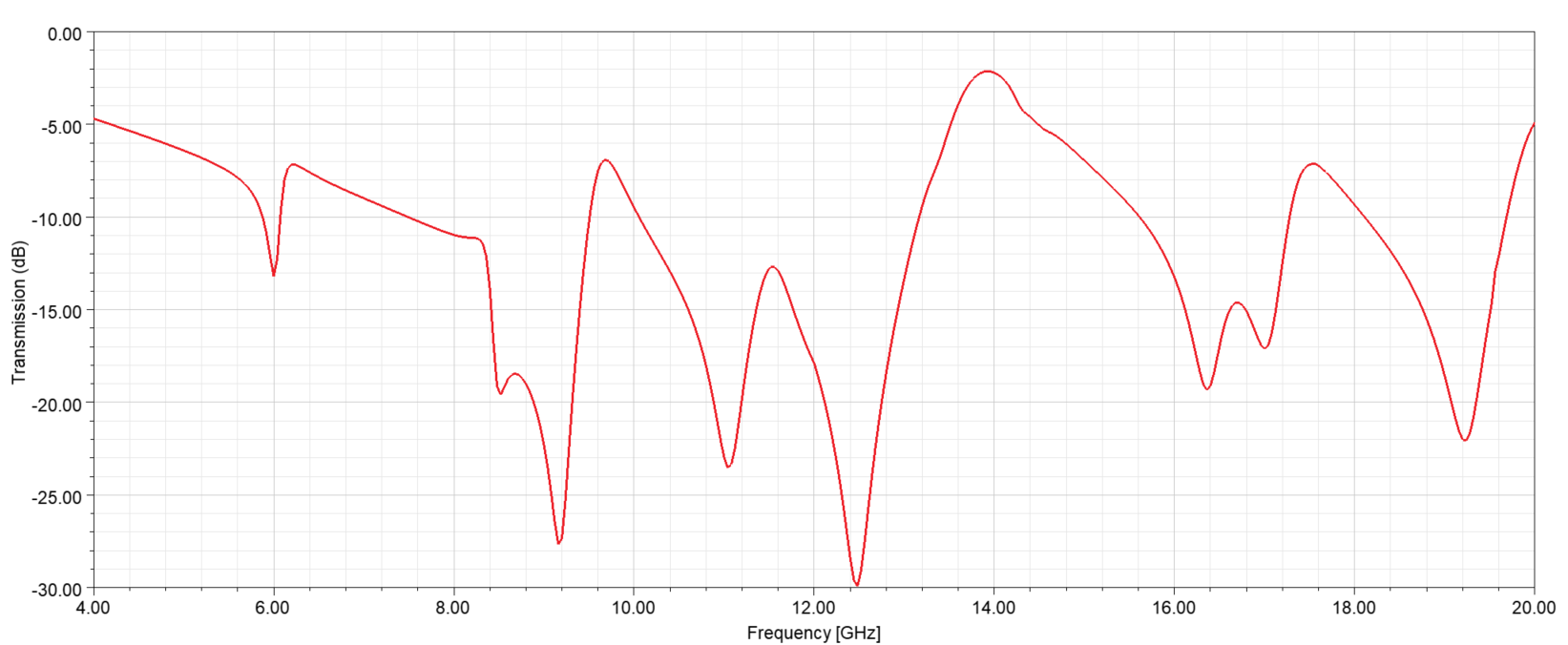}
    \caption{Transmission coefficient of two-layer FSS with 0 mm offset of the top pattern}
\end{figure}

The top FSS unit cell was then evaluated at offsets of 2 mm, 3 mm, and finally a full 5 mm from the lower layer.  Results obtained for transmission and reflection characteristics are shown in Figures 7 and 8 respectively.

\begin{figure}[h]
\center
   \includegraphics[width=\linewidth]{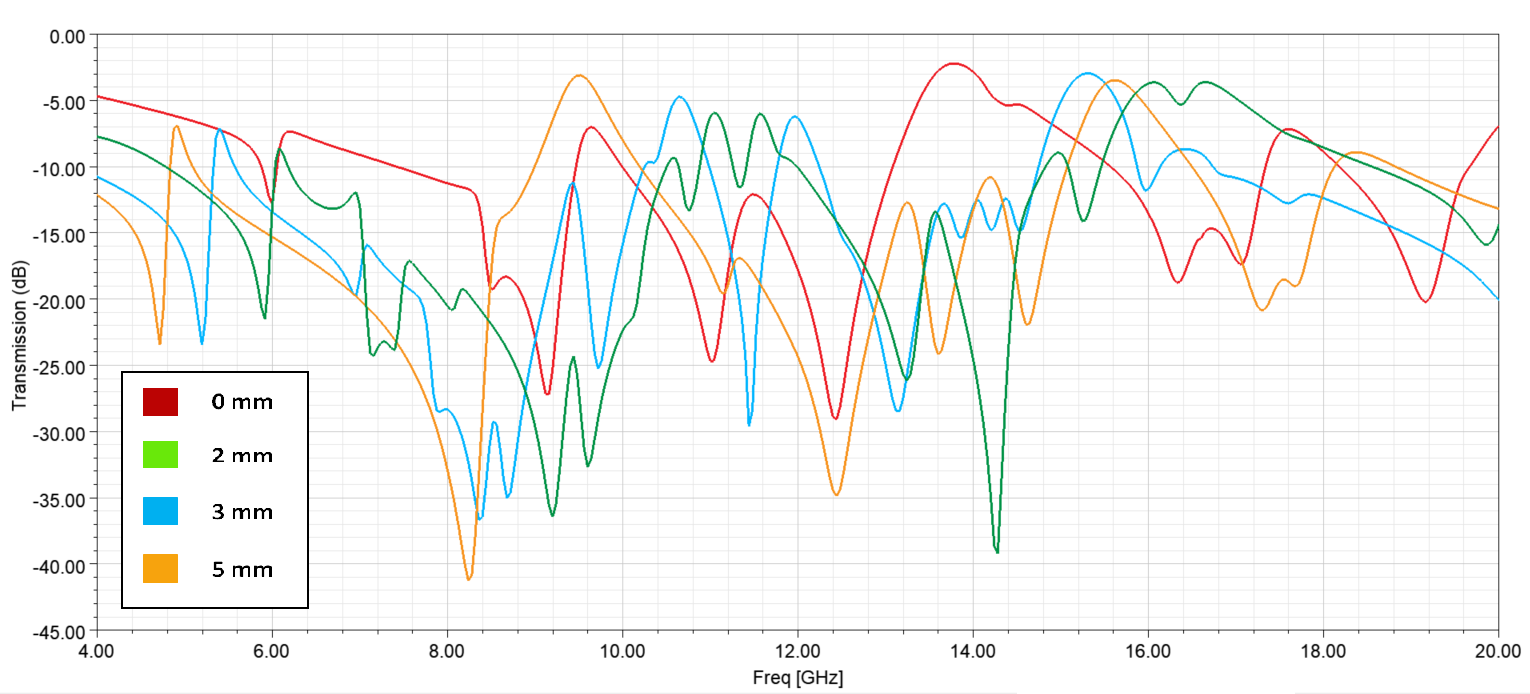}
    \caption{Transmission coefficient of two-layer FSS with varying pattern offsets}
\end{figure}

\begin{figure}[h]
\center
   \includegraphics[width=\linewidth]{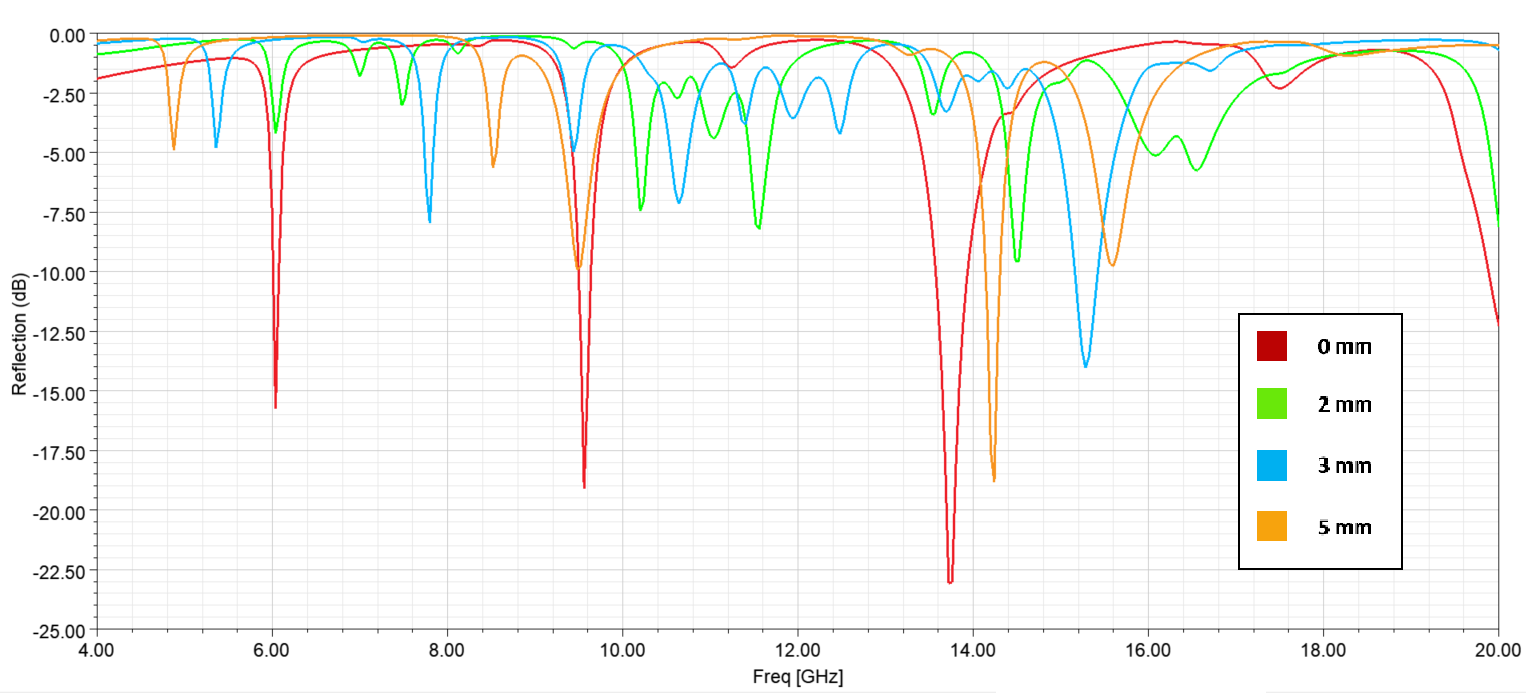}
    \caption{Reflection coefficient of two-layer FSS with varying pattern offset}
\end{figure}

\break

Results obtained generally indicate bandpass performance is best in the case of no offset of the top layer with the greatest observed reflection attenuation occurring in this case.  Bandstop structure resonances varied considerably in each of the cases, though increasing offset generally led to higher attenuation of transmission characteristics and somewhat wider bandwidths.  Top layer substrate thickness was also varied parametrically in the zero and 5 mm cases to determine the impact of varying the top layer thickness dimension.  The transmission and reflection parameters for the zero offset case are shown in Figs. 9 and 10.

\begin{figure}[h]
\center
   \includegraphics[width=\linewidth]{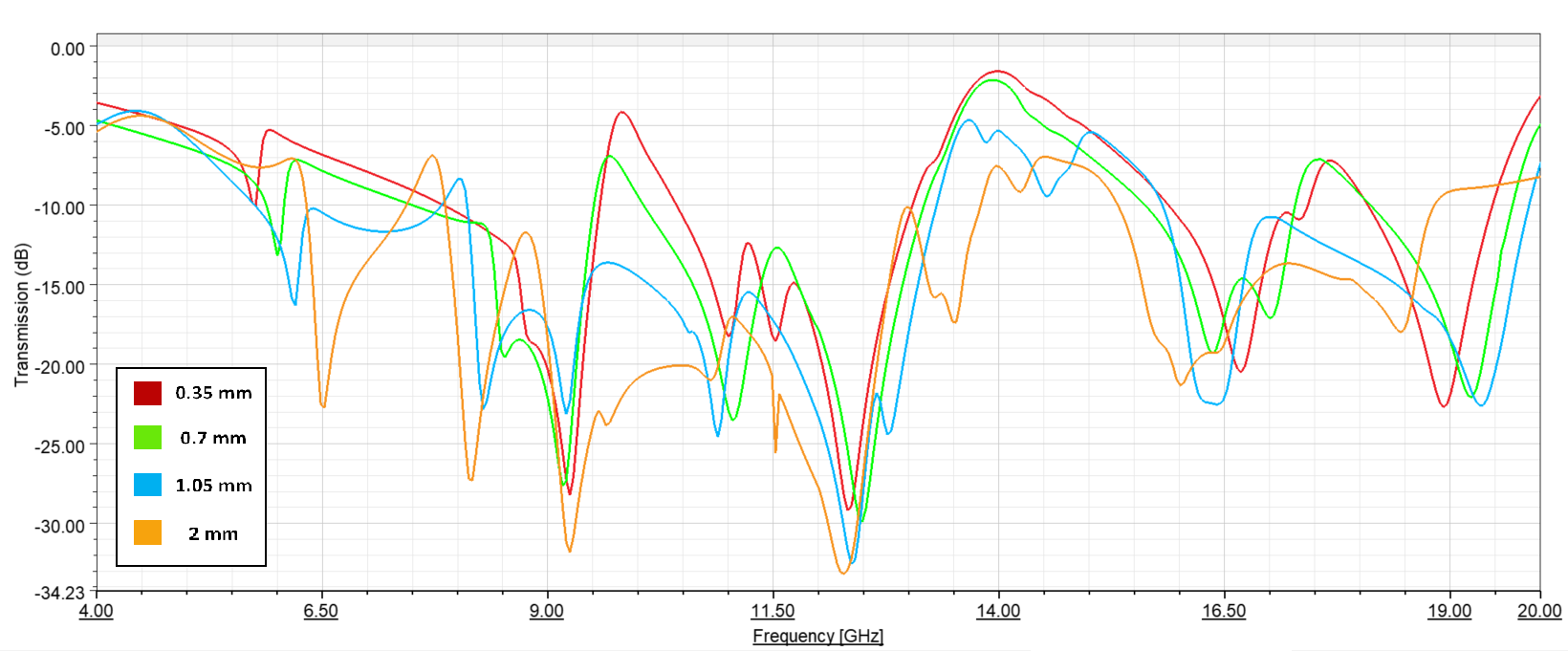}
    \caption{Transmission coefficient of two-layer FSS with 0 mm pattern offset and varying thickness}
\end{figure}

\begin{figure}[h]
\center
   \includegraphics[width=\linewidth]{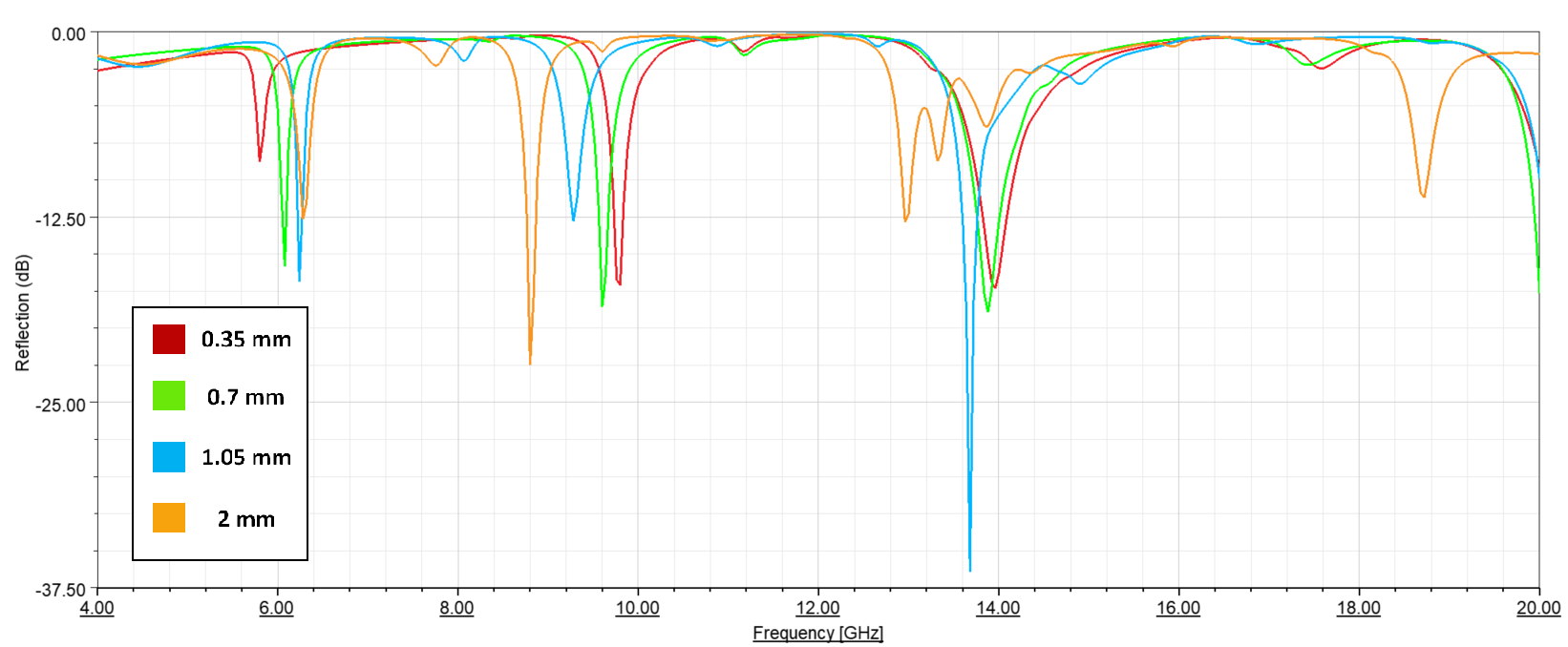}
    \caption{Reflection coefficient of two-layer FSS with 0 mm pattern offset and varying thickness}
\end{figure}

Bandpass performance for the varying thicknesses clustered around approximately 6.1 GHz, 9.3 GHz, and 13.6 GHz with substrate thicknesses variations causing shifting either to higher or lower frequencies.  Bandstop characteristics generally followed the same trend as thickness was varied with the same large-scale features apparent in the transmission parameter, though the best bandstop performance was noted in the thickest case with the top layer substrate at 2 mm.  Here five bandstop resonances with transmission at -20 dB or lower were observed at 6.5 GHz, 8.5 GHz, 9.2 GHz, 12.3 GHz, and 16.0 GHz.  In the case where the top pattern was shifted the full 5 mm, results for transmission and reflection are illustrated in Figs. 11 and 12.  Bandpass effects were still evident around 14.0 GHz and below 9.8 GHz though performance was lessened in this case.  In both the 0 mm and 5 mm offset cases minimum reflection coefficient values seem likely to arise with a top substrate thickness between 1.05 mm and 2.0 mm suggesting an optimum for bandpass design.  The bandstop performance of the FSS stack was much more variable than with a top pattern offset of zero, indicating the sensitivity of bandstop resonances to substrate thicknesses.  Generally, an increasing dielectric thickness corresponded with decreased bandstop performance, with the maximum transmission attenuation occurring in the 0.35 mm thickness instance.  Five structure resonances were again noted for this thickness, though slightly shifted from the previous resonances in the zero mm pattern offset stack, and occurring at 6.1 GHz, 9.6 GHz, 11.4 GHz, 14.8 GHz, and 17.9 GHz.  Several cases of further increased substrate thickness were parametrically explored and transmission and reflection data is shown in Figs. 13 and 14.  

\begin{figure}[h]
\center
   \includegraphics[width=\linewidth]{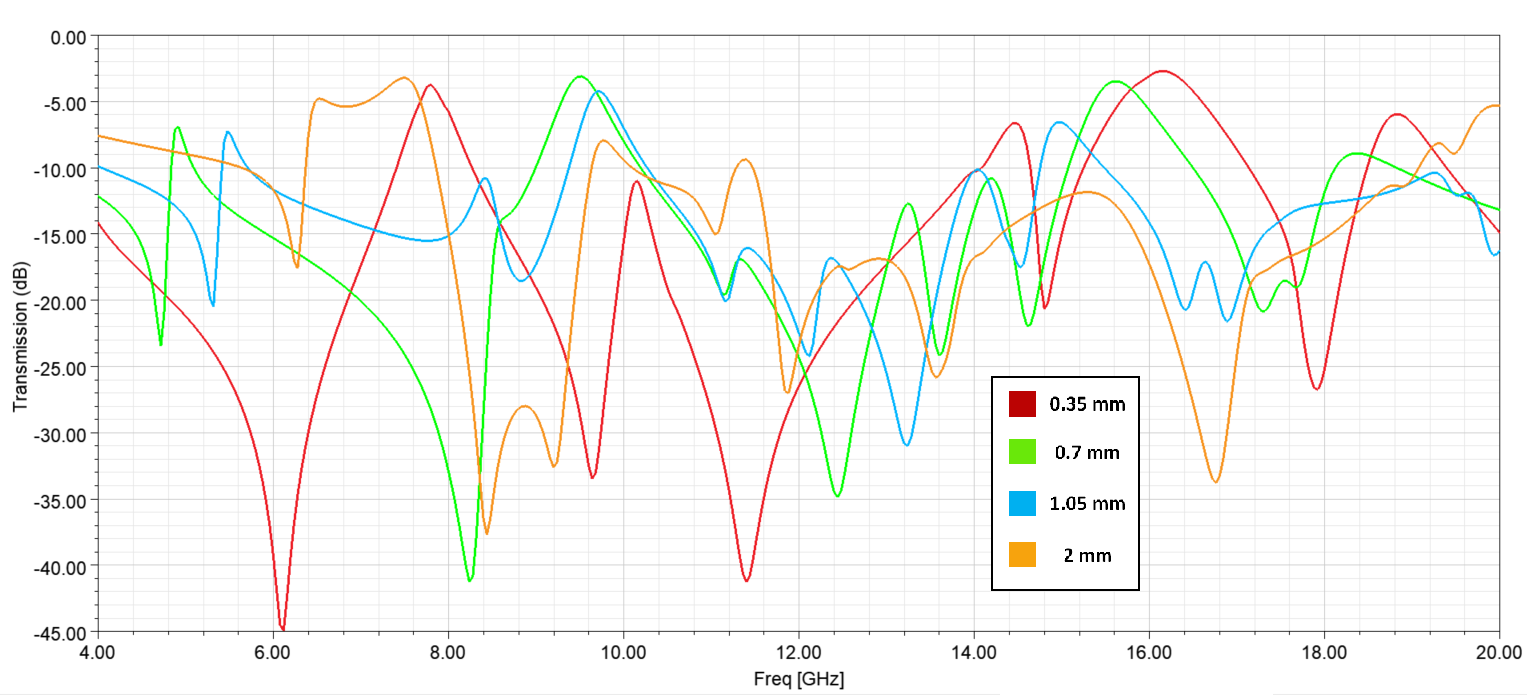}
    \caption{Transmission coefficient of two-layer FSS with 5 mm pattern offset and varying thickness}
\end{figure}

\begin{figure}[h]
\center
   \includegraphics[width=\linewidth]{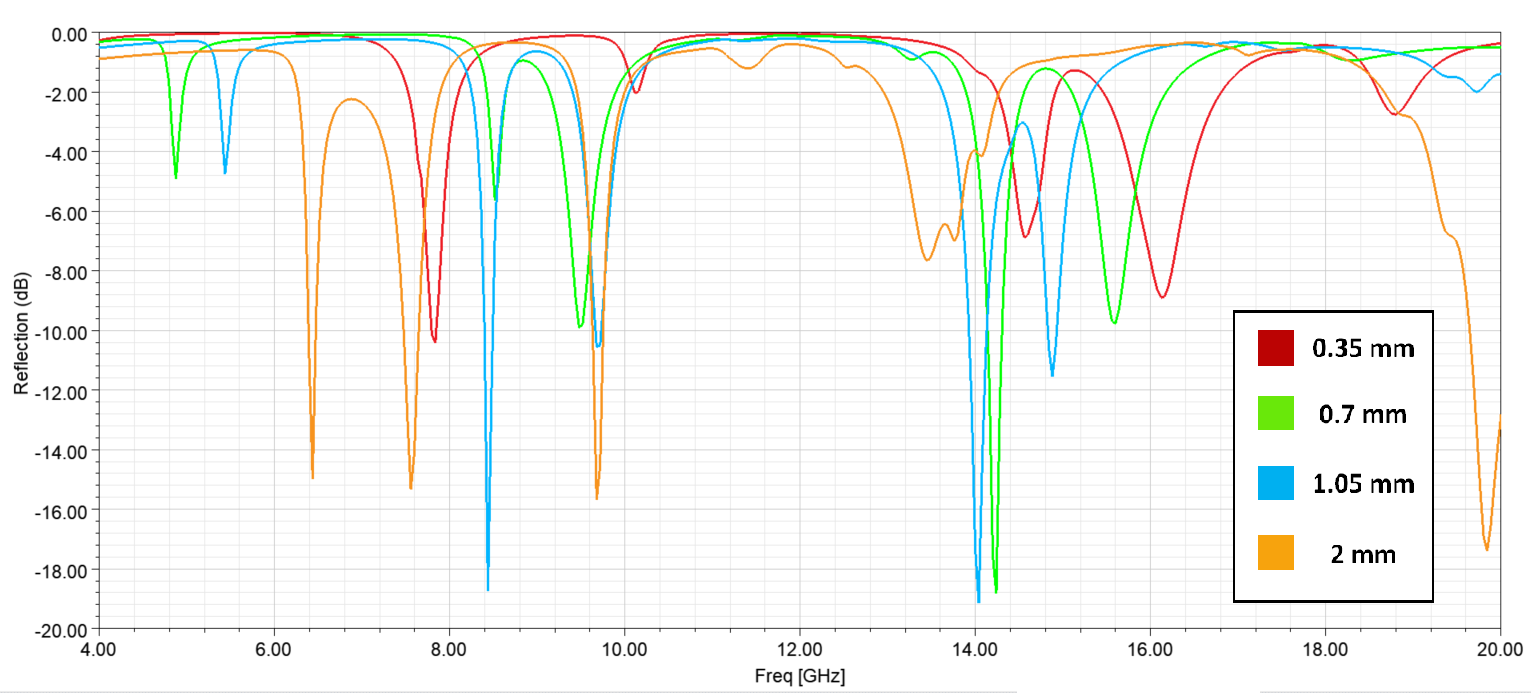}
    \caption{Reflection coefficient of two-layer FSS with 5 mm pattern offset and varying thickness}
\end{figure}

\begin{figure}[hbt]
\center
   \includegraphics[width=\linewidth]{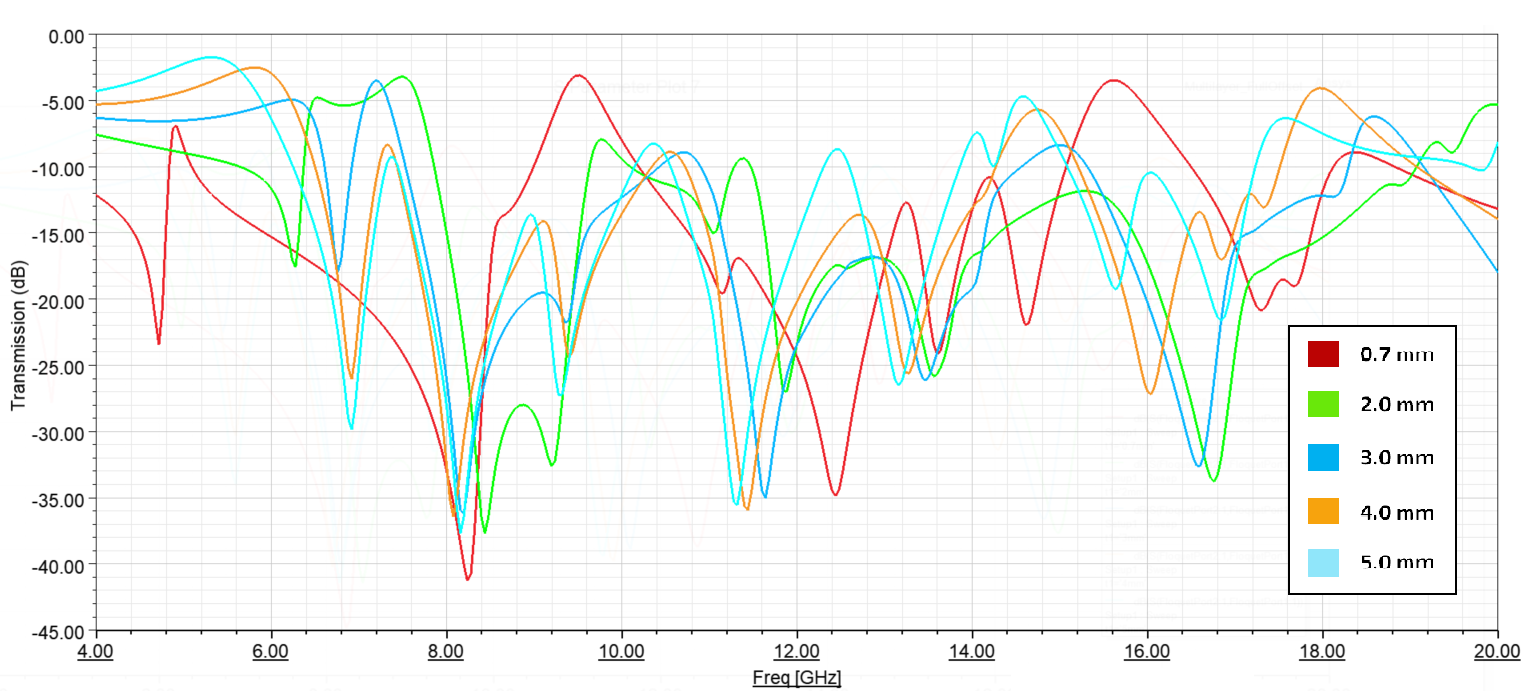}
    \caption{Transmission coefficient of two-layer FSS with 5 mm pattern offset and increased varying thickness}
\end{figure}

\begin{figure}[hbt]
\center
   \includegraphics[width=\linewidth]{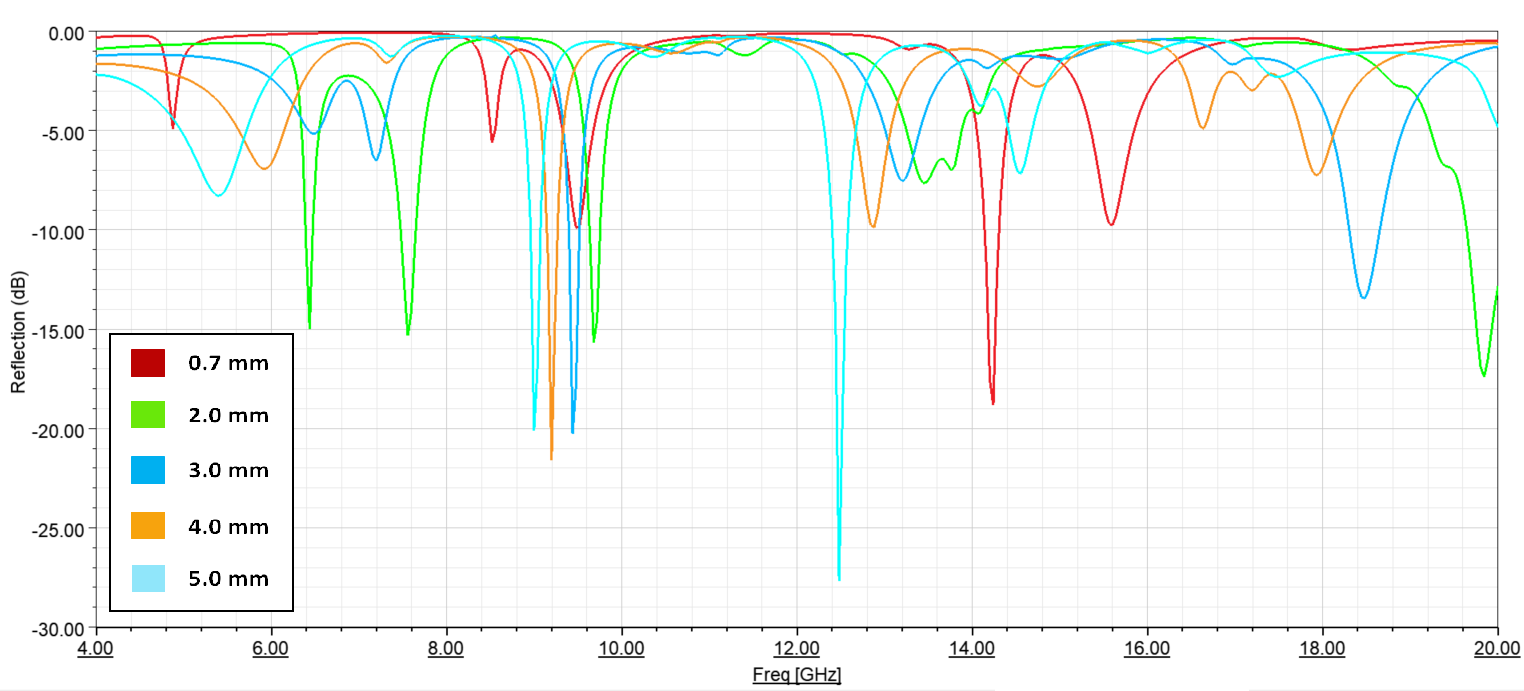}
    \caption{Reflection coefficient of two-layer FSS with 5 mm pattern offset and increased varying thickness}
\end{figure}

\hfill \break
\hfill \break
\hfill \break
\hfill \break
\hfill \break
\hfill \break
\hfill \break

\section{Third Layer Addition}

After completing parametric evaluations of the two-layer FSS structure, a third layer was added to the simulation to evaluate performance of a larger stack.  The unit cell morphology was identical to the previous two layers, and the entire structure is shown in Fig. 15. 

\begin{figure}[hbt]
\center
   \includegraphics[width=\linewidth]{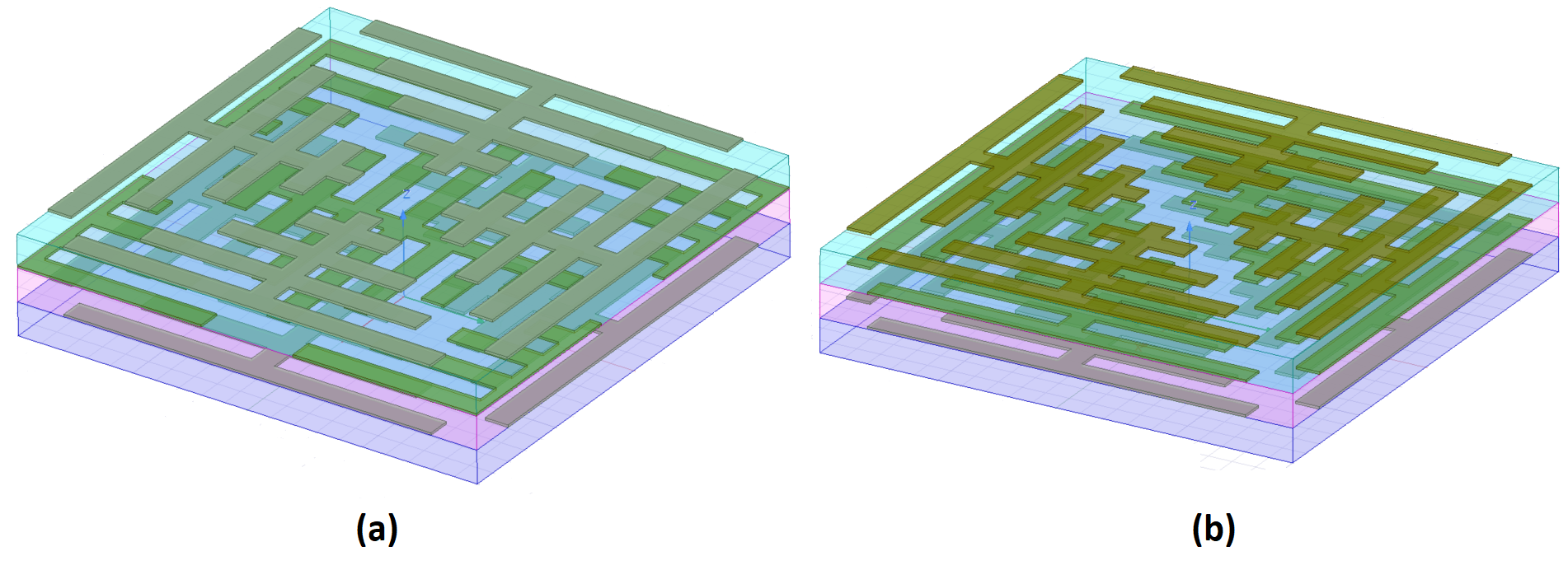}
    \caption{Three-layer FSS with a) 5 mm (max) offset of the middle layer pattern and b) 0 mm offset}
\end{figure}

Two cases were simulated, the first consisted of a middle layer pattern offset of 5 mm, while in the second the unit cell pattern was fully aligned for all three layers.  Reflection data for the three-layer structure is shown in Figs. 16 and 17 and transmission data is illustrated in Fig. 18 and 19.  The lack of pattern offset reduces bandpass characteristics as the number of low-reflection notches decreases, though the bandpass windows at approximately 6.3 GHz, 13.6 GHz, and 19.2 GHZ are deepened and slightly shifted towards higher frequencies in the zero-offset case.  Bandstop characteristics show marked differences in both cases, though maximum transmission attenuation is seen in the 5 mm offset case.  

\begin{figure}[htbp]
\center
   \includegraphics[width=\linewidth]{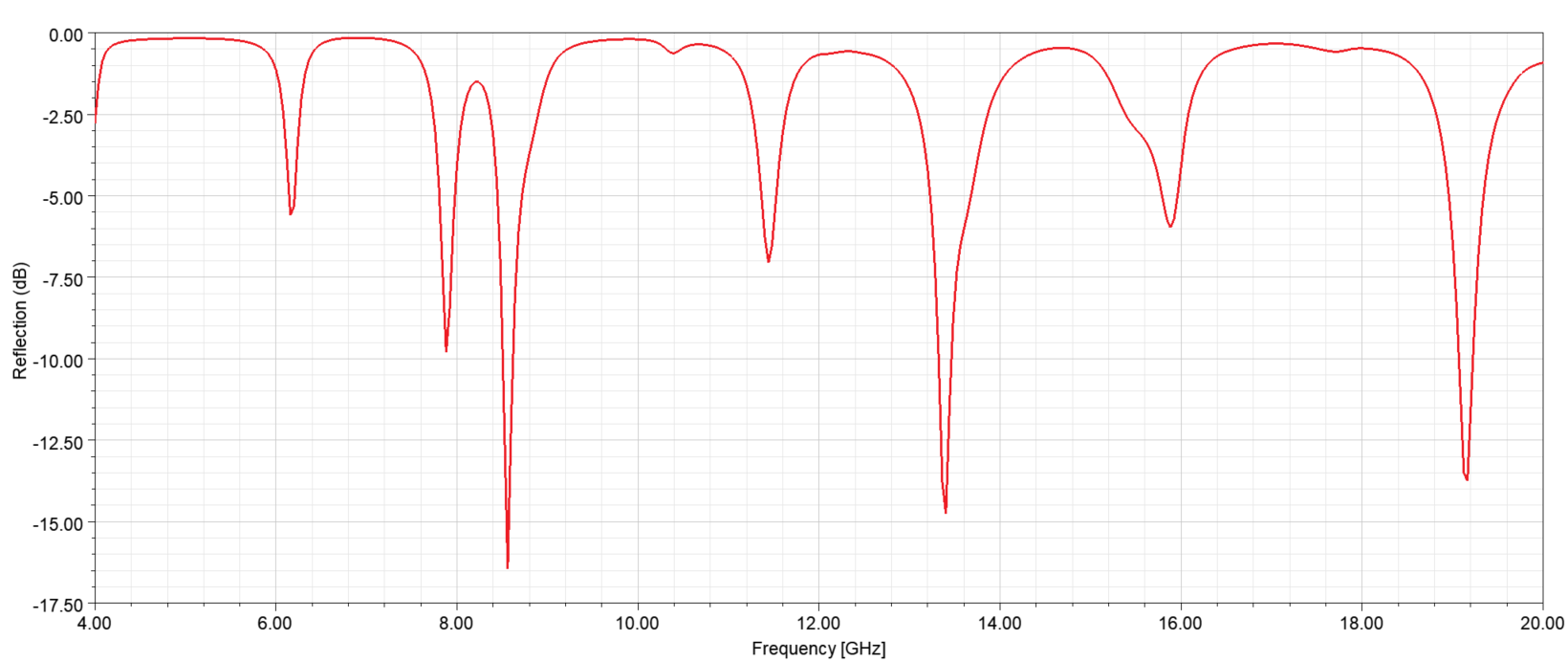}
    \caption{Reflection coefficient of three-layer FSS with 5 mm pattern offset}
\end{figure}

\begin{figure}
\center
   \includegraphics[width=\linewidth]{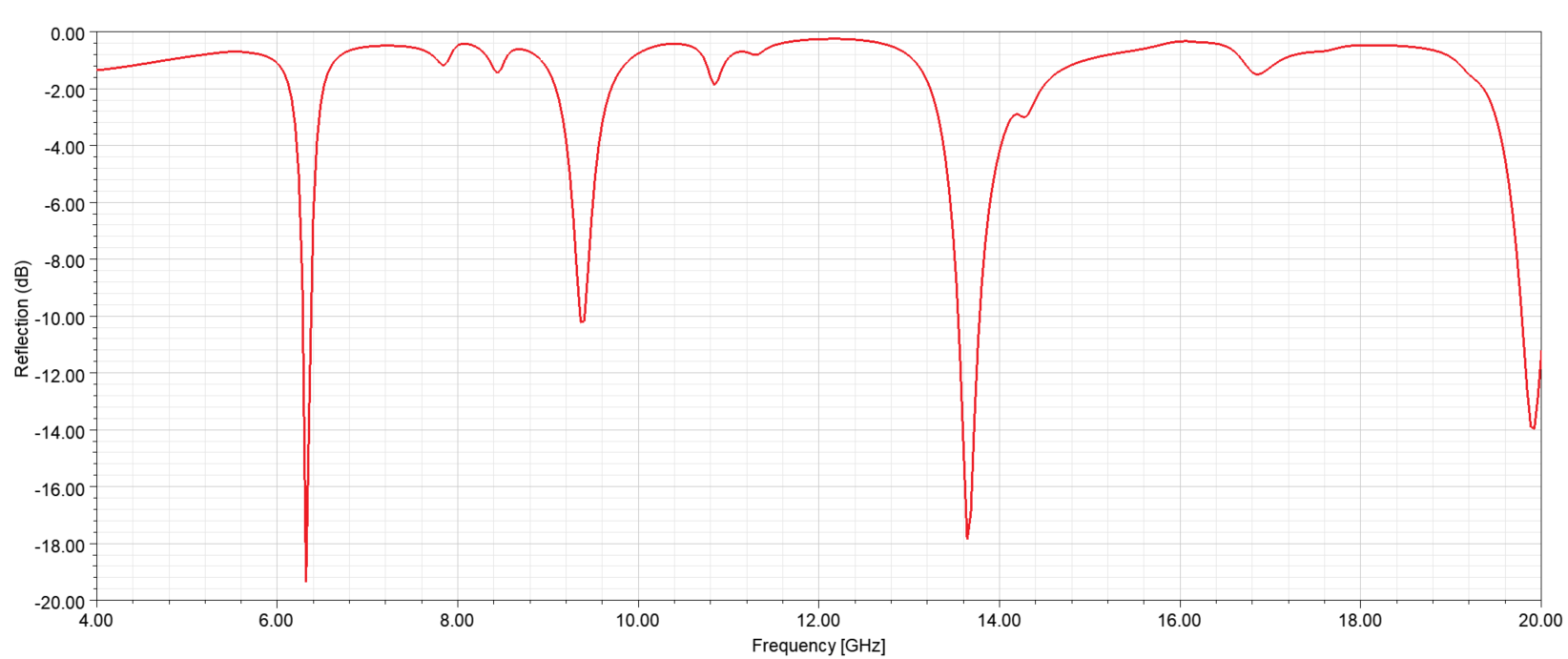}
    \caption{Reflection coefficient of three-layer FSS with 0 mm pattern offset}
\end{figure}

\begin{figure}
\center
   \includegraphics[width=\linewidth]{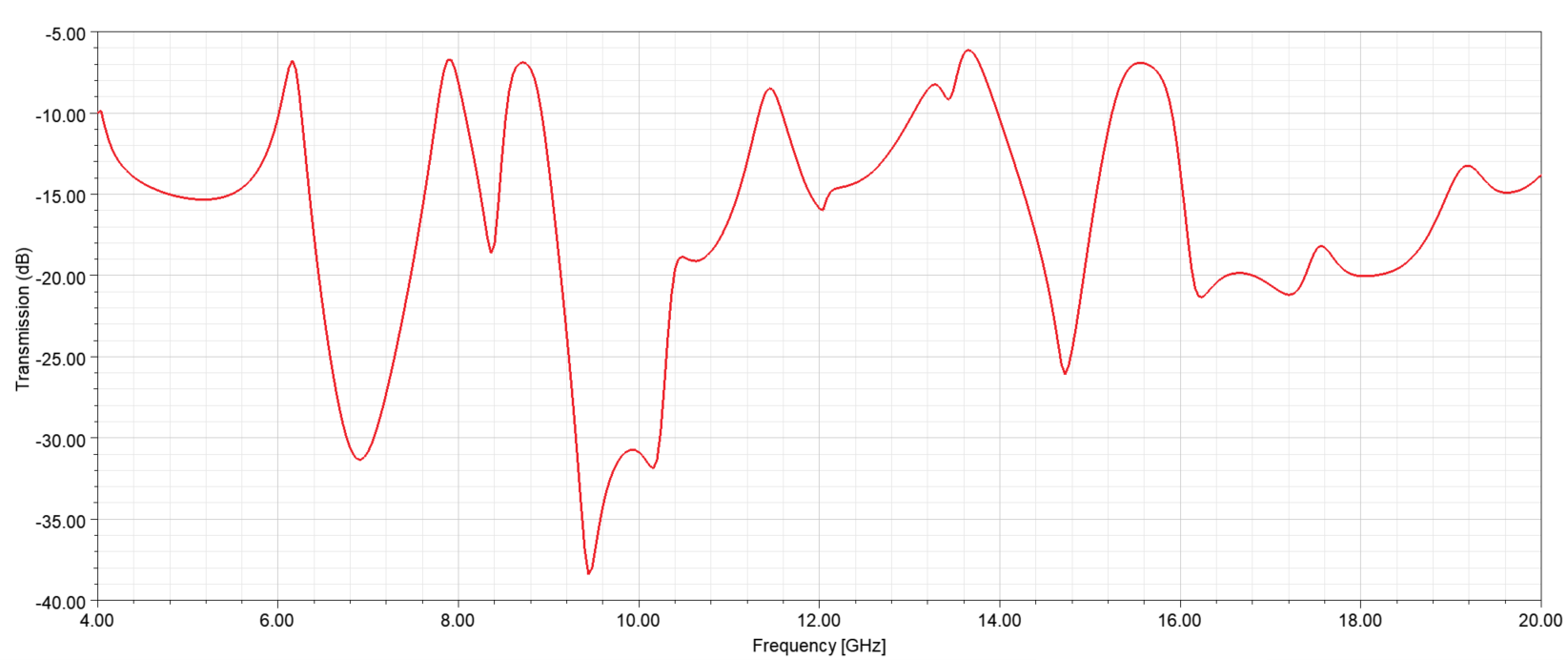}
    \caption{Transmission coefficient of three-layer FSS with 5 mm pattern offset}
\end{figure}

\begin{figure}
\center
   \includegraphics[width=\linewidth]{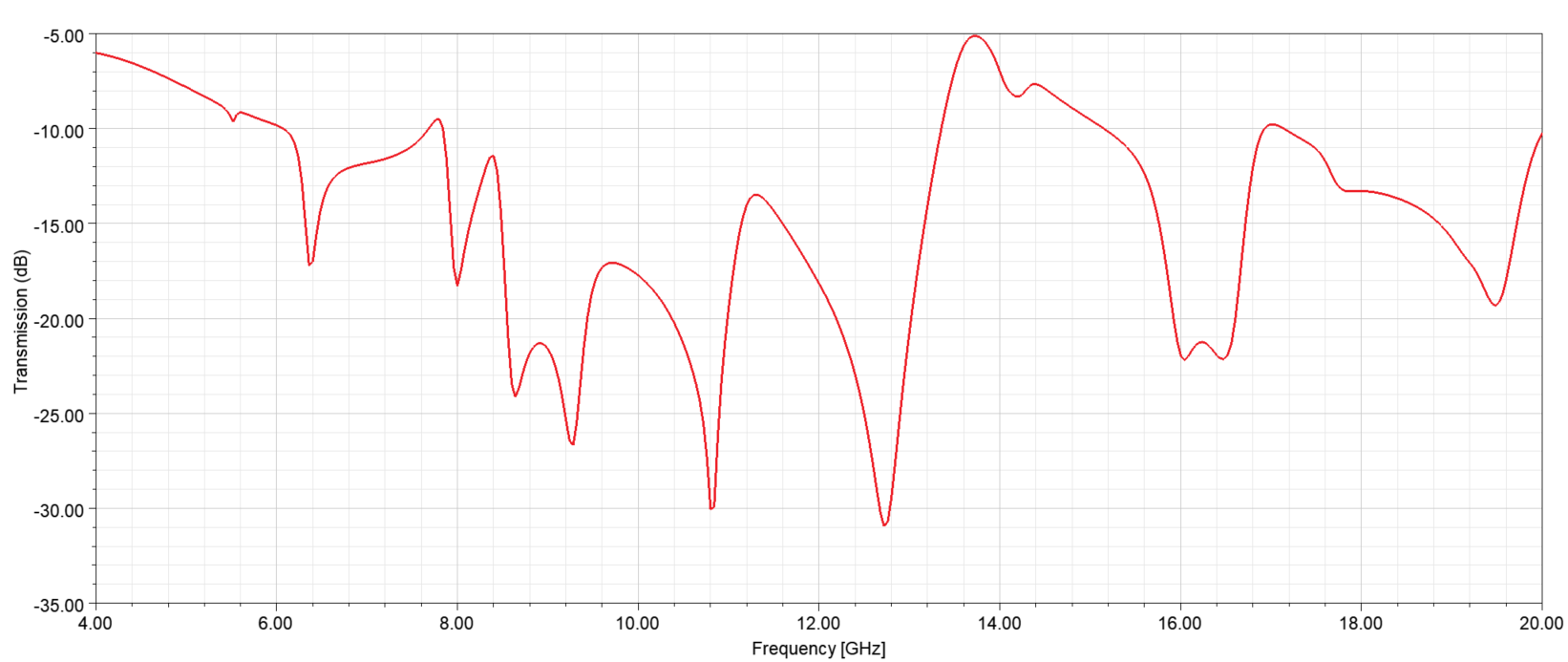}
    \caption{Transmission coefficient of three-layer FSS with 0 mm pattern offset}
\end{figure}

\section{Conclusions}

Numerical simulations of fractal-based unit cell morphologies for FSS structures indicate several design parameters that may be varied to achieve different results depending on desired outcomes.  The three factors studied were number of FSS layers in the total structure, thickness of the layers, and pattern offset.  Generally, simply varying one parameter does not guarantee desired results, as both bandpass and bandstop characteristics are shown to be dependent on the interactions of multiple variables.  Broadly an increase in the number of layers of the FSS stack creates more bandpass and bandstop features, though this is in part dependent on the offset of the pattern on one layer with regards to that of a different layer.  Greatest reflection coefficient attenuation was seen in the case of no relative offset between patterns, while highest transmission coefficient attenuation was observed in the case of maximum pattern offset, thus any design to maximize attenuation of either reflection or transmission would likely yield a design tradeoff between the two.  In the case of no offset between bottom and top patterns, increasing top layer thickness principally increased attenuation of the transmission coefficient though this was not the case in the instance of maximum pattern offset.  Reflection coefficients varied with thickness in both the zero and maximum offset cases.  In the future, further work may focus on additional parameter variations as well as examination of other variables and their influence, including pattern dimensions, pattern orientations, substrate materials, or other fractal-type pattern designs.


\begin{thebibliography}{99}
\bibitem{1} B. A. Munk, Finite Antenna Arrays and FSS, Hoboken, NJ, USA: Wiley
2003.

\bibitem{2} Rahim, Tariq. (2015). X-band Band-pass Frequency Selective Surface for Radome Applications. TELKOMNIKA Indonesian Journal of Electrical Engineering. 16. 10.11591/tijee.v16i2.1613.

\bibitem{3} H. Paik and S. Paik, "A Compact Frequency Selective Surface based X band Spatial Bandpass Filter Realization," 2022 3rd International Conference for Emerging Technology (INCET), 2022, pp. 1-4, doi: 10.1109/INCET54531.2022.9824774.

\bibitem{4} W.-T. Wang, S.-X. Gong, X. Wang, H.-W. Yuan, J. Ling \& T.-T. Wan (2009) RCS Reduction of Array Antenna by Using Bandstop FSS Reflector, Journal of Electromagnetic Waves and Applications, 23:11-12, 1505-1514, DOI: 10.1163/156939309789476473.

\bibitem{5} M. Yan, J. Wang, S. Qu, M. Feng, Z. Li, H. Chen, J. Zhang \& L. Zheng (2016) Highly-selective, closely-spaced, dual-band FSS with second-order characteristic, IET Microwaves, Antennas, \& Propagation, Vol. 10 Issue 10, pp. 1087-1091.

\bibitem{6} L. Kurra, M. P. Abegaonkar, A. Basu and S. K. Koul, "FSS Properties of a Uniplanar EBG and Its Application in Directivity Enhancement of a Microstrip Antenna," in IEEE Antennas and Wireless Propagation Letters, vol. 15, pp. 1606-1609, 2016, doi: 10.1109/LAWP.2016.2518299.

\bibitem{7} T. Rahim, F. A. Khan and X. Jiadong, "Design of X-band frequency selective surface (FSS) with band pass characteristics based on miniaturized unit cell," 2016 13th International Bhurban Conference on Applied Sciences and Technology (IBCAST), 2016, pp. 592-594, doi: 10.1109/IBCAST.2016.7429937.

\bibitem{8} Sarika, R. Kumar, M. R. Tripathy and D. Ronnow, "Fractal frequency selective surface based band stop filters for X-band and Ku-band applications," 2017 3rd International Conference on Advances in Computing,Communication \& Automation (ICACCA) (Fall), 2017, pp. 1-4, doi: 10.1109/ICACCAF.2017.8344692.


\end{thebibliography}
\end{document}